\documentstyle[twoside,fleqn,espcrc2,draft]{article}

\newcommand{\AmS}{{\protect\the\textfont2
  A\kern-.1667em\lower.5ex\hbox{M}\kern-.125emS}}
\newcommand{\ssc}{\scriptscriptstyle}

\newcommand{\scite}{~\cite}
\newcommand{\quattrova}{($\varphi^{\scriptscriptstyle a},\,
c^{\scriptscriptstyle a},\,
\lambda_{\scriptscriptstyle  a},\,
{\bar c}_{\scriptscriptstyle a}$)}  %$
\newcommand{\be}{\begin{equation}}
\newcommand{\ee}{\end{equation}}
\newcommand{\bea}{\begin{eqnarray}}
\newcommand{\eea}{\end{eqnarray}}

\title{Functional Techniques in Classical Mechanics}
\author{Ennio Gozzi\address{Department of Theoretical Physics,
University of Trieste \\
Strada Costiera 11, Miramare-Grignano 34014, Trieste\\
and INFN, Trieste, Italy.}}

\begin{document}

\begin{abstract}
In 1931 Koopman and von Neumann extended previous work of Liouville
and provided an operatorial version of Classical Mechanics (CM).
In this talk we will review a path-integral formulation of this operatorial
version of CM. In particular we will study the geometrical nature of the many
auxiliary variables present and of the unexpected universal symmetries 
generated by the functional technique.
\end{abstract}

\maketitle

\section{INTRODUCTION}
I usually do not go  to conferences which have in the title the standard
fashionable words "{\it Quantum Gravity}" (QG). I made an exception this time
because my dear friend
Giampiero had the good taste of creating two sessions: one dedicated to QG
and another to "{\it Foundations of Quantization}"(FQ). In fact I belong 
to that minority which thinks that we should not
only try to attack the second of the two horns of the problem of 
{\it Quantum Gravity}, that is {\it Gravity}, but also the first one that is 
the {\it Quantum}. Gravity is the queen of {\it geometrical} theories, and I
feel  we should better
understand QM from a more {\it geometrical} point of view before putting
the two theories together.
Attempts in this direction already exist  like for example the method called
 "{\it Geometric Quantization}"(see ref.\scite{wood1} for a review).
We feel anyhow that method should be made
less cumbersome and understood from a more physical point of view. 
With this goal in mind we proved\scite{lesha} that, 
by formulating classical mechanics (CM) via functional methods\scite{enniocl},
the standard geometric quantization  rules become equivalent 
to freezing to zero some Grassmannian partners of time. This may throw
some light on the geometrical aspects of quantization and work is in progress on
it\scite{lesha}. 
In this paper we will limit ourselves to reviewing the classical mechanics
part of this project that means the geometrical structures which enter
the functional approach to CM.

\section{FUNCTIONAL APPROACH TO CLASSICAL MECHANICS}

The functional  formulation of CM mentioned above 
is a path-integral  approach to the
{\it operatorial\/} version of CM proposed by Koopman and von
Neumann\scite{koop} in 1931. These authors, instead of using the Hamiltonian
and the Poisson brackets for the classical evolution of a system, 
used the well-known Liouville operator,\break
$\displaystyle {\hat L}\equiv{\partial H\over\partial p} {\partial\over\partial q}-
{\partial H\over\partial q}{\partial\over\partial p},$
and the associated commutators. One can generalize their formalism 
to higher forms and get what is known as
the Lie derivative of the Hamiltonian flow, \scite{marsd}.  It was shown in
\scite{enniocl} that the operatorial formalism mentioned above  could
have a functional or path integral counterpart. The procedure goes as follows.
Let us denote by ${\cal M}$ our phase space with $2n$ phase-space coordinates
$\varphi^{a}=(q^{\ssc j},p^{\ssc j})$ (the index "$a$" spans both $q$'s and
$p$'s) and by $H(\varphi)$ the Hamiltonian of the system. 
The classical  trajectories are solutions of the Hamilton equations of motion:
${\dot \varphi^{a}}=\omega^{ab}{\partial H\over \partial\phi^{b}}$ where
$\omega^{ab}$ is the standard symplectic matrix. 
A natural object to introduce is  the {\it classical} analog,~
$Z_{\scriptscriptstyle CM}[j]$,  of the quantum generating 
functional:

\be
\label{eq:due}
Z_{\scriptscriptstyle CM}[j]=N\int{\cal D}\varphi{\tilde{\delta}}
[\varphi(t)-\varphi_{cl}(t)]\exp\int j\varphi dt
\end{equation}  

\noindent where $\varphi$ are the $\varphi^{a}\in{\cal M}$, $\varphi_{cl}$ are 
the solutions of the equations of motion,
$j$ is an external current and $\widetilde{\delta}[\;\;]$ is a functional 
Dirac-delta which forces
every path $\varphi(t)$ to sit on a classical one $\varphi_{cl}(t)$.  
There are all
possible initial conditions integrated over in (\ref{eq:due}) and, because of this,
one should be very careful in properly defining the measure 
of integration and the functional
Dirac delta.  We should now check whether the path integral of eq.~(\ref{eq:due}) leads to the
well 
known operatorial 
formulation\scite{koop} of CM . To do that let us first
rewrite the functional Dirac delta in (\ref{eq:due}) as:
\bea
\label{eq:tre}
\lefteqn{{\tilde\delta}[\varphi(t)-\varphi _{cl}(t)]=}\\
\lefteqn{\qquad={\tilde\delta}[{\dot\varphi 
^{a}-\omega^{ab}
\partial_{b}H]\vert\det [\delta^{a}_{b}\partial_{t}-\omega^{ac}\partial_{c}\partial
_{b}H}]\vert}\nonumber
\eea 
\noindent The determinant which appears in (\ref{eq:tre}) is always positive and
so we can drop the modulus sign $|\;\;|$. 
The next step is to insert (\ref{eq:tre}) in (\ref{eq:due}) and write the 
$\tilde{\delta}[\;\;]$
as a Fourier transform over some new variables $\lambda_{a}$, i.e.:
\bea
\label{eq:quattro}
{\tilde{\delta}}\biggl[{\dot\varphi }^{a}&-&\omega^{ab}{\partial 
H\over\partial\varphi ^{b}}\biggr]=\\
&=& \int~{\cal D}\lambda_{a}~\exp~i\int\lambda_{a}\biggl[{\dot \varphi 
}^{a}-\omega^{ab}
{\partial H\over\partial\varphi ^{b}}\biggr]dt\nonumber
\eea 
Next we express the determinant in eq.(\ref{eq:tre}) via
grassmannian variables ${\bar c}_a, c^{a}$:
\be
\label{eq:cinque}
\int{\cal D}c^{a}{\cal D}{\bar c}_{a}\exp\biggl[-\int {\bar c}_{a}
[\delta^{a}_{b}\partial_{t}-\omega^{ac}\partial_{c}
\partial_{b}H]c^{b}~dt\biggr]
\ee 
\noindent Inserting (\ref{eq:tre}),(\ref{eq:quattro}) and (\ref{eq:cinque})  
in (\ref{eq:due}) we get:
\begin{equation}
\label{eq:sei}
Z_{\scriptscriptstyle CM}[0]=\int{\cal D}\varphi^{a}{\cal D}\lambda_{a}{\cal 
D}c^{a}{\cal D}{\bar c}_{a}\exp\biggl[i\int~dt{\widetilde{\cal L}}\biggr]
\end{equation}  
\noindent where $\widetilde{\cal L}$ is:
\begin{equation}
\label{eq:sette}
{\widetilde{\cal L}}=\lambda_{a}[{\dot\varphi }^{a}-\omega^{ab}\partial_{b}H]+
i{\bar c}_{a}[\delta^{a}_{b}\partial_{t}-\omega^{ac}\partial_{c}\partial_{b}H]
c^{b}
\end{equation}  
\noindent Varying this Lagrangian with respect to
$\lambda_{a}$, one can easily obtain  the standard equation
of motion for $\varphi^{a}$, while varying it with respect to ${\bar c}_{a}$ 
gives the
 equations for the $c^{a}$. The overall set of equations of motion is:
\bea
\label{eq:otto}
{\dot\varphi}^{a}-\omega^{ab}\partial_{b}H & = & 0\nonumber\\
\bigl[\delta^{a}_{b}\partial_{t}-\omega^{ac}\partial_{c}
\partial_{b}H\bigr]c^{b} & = & 0\\
\delta^{a}_{b}\partial_{t}{\bar c}_{a}+{\bar c}_{a}\omega^{ac}\partial_{c}
\partial_{b}H & = & 0\nonumber\\
\bigl[\delta_{b}^{a}\partial_{t}+\omega^{ac}\partial_{c}\partial_{b}
H\bigr]\lambda_{a} & = & -i{\bar c}_{a}\omega^{ac}\partial_{c}
\partial_{d}\partial_{b}H c^{d}\nonumber
\eea  
 The Hamiltonian associated to 
the $\widetilde{\cal L}$ of eq.(\ref{eq:sette})is:
\begin{equation}
\label{eq:nove}
\widetilde{\cal H}=\lambda_a\omega^{ab}\partial_bH+i\bar{c}_a\omega^{ac}
(\partial_c\partial_bH)c^{b}
\end{equation} 
The equations of motion of eqs.(\ref{eq:otto}) can be obtained also from this
Hamiltonian using some
extended Poisson brackets ({\it EPB}) defined in the space \quattrova
~as follows:
$ \{\varphi ^{a},\lambda_{b}\}_{\scriptscriptstyle EPB}=\delta^{a}_{b}~~;
~~\{{\bar c}_{b}, c^{a}\}_{\scriptscriptstyle EPB}=-i\delta^{a}_{b}$.
All the other brackets are zero. 
Having a path-integral we can also define
the concept of commutators\scite{enniocl} and realize the various variables as operators.
It is then easy to prove\scite{enniocl} that the bosonic part of $\widetilde{\cal H}$
turns into the Liouville operator of Koopman and von Neumann. This confirms
that our path-integral (\ref{eq:due}) is the correct counterpart of the
operatorial approach to CM\scite{koop}. We skip here these details which can
be found in the first paper of ref.\scite{enniocl}.
The reader may not like the pletora of variables \quattrova  that we had to
introduce. It is actually possible to simplify things considerably by introducing
the concept of superfield.
If we enlarge the base space, which is up to now just the time t, by including 
two grassmannian partners of time,
$\theta$ and $\bar{\theta}$, we can put 
together all the 
variables \quattrova~ into the following superfield:
\be
\Phi^a=\varphi^a+\theta
c^a+\bar{\theta}\omega^{ab}\bar{c}_b+i\bar{\theta}\theta\omega^{ab}\lambda_b
\ee
Via this superfield the complicated expression of ${\cal\widetilde{H}}$ can
be written as
${\cal \widetilde{H}}=i\int d\theta d\bar{\theta}H[\Phi]$. This formula
is the starting point for the quantization procedure we presented in ref.\scite
{lesha}. Besides this unification , we shall also show in the next 
section that all the variables  \quattrova have a clear geometrical meaning and they are all
needed to shed light on our construction.  
 
\section{GEOMETRICAL CONTENT}

>From the equations of motion (\ref{eq:otto})
one  notice immediately that $c^b$ transforms under 
the Hamiltonian vector 
field\scite{marsd} $h\equiv\omega^{ab}\partial_bH\partial_{a}$
as  the {\it basis} $d\varphi^{b}$ 
 of generic forms $\alpha\equiv \alpha_{a}(\varphi)d\varphi^{a}$
. It is possible to show that this  happens not only under the Hamiltonian
flow but also 
under any diffeomorphism of the phase-space ${\cal M}$ whose coordinates are 
${\varphi^a}$. We can look at the $c^{a}$ in a manner dual to the
previous one because we can say that  the $c^{a}$ also transform as
{\it components} of  tangent vectors:
$V^{a}(\varphi){\partial\over\partial\phi^{a}}$.
Because of this the space whose coordinates are $(\varphi^{a}, c^{a})$ is called in \scite{Schw} 
{\it reversed-parity tangent bundle} and is indicated as $\Pi T{\cal M}$. The 
"{\it reversed-parity}" specification
is because the $c^{a}$ are Grassmannian variables. From the Lagrangian
(\ref{eq:sette}) we notice that the 
$(\lambda_{a},{\bar c}_{a})$ are the "momenta" of the  variables 
$(\varphi^{a},c^{a})$, so
we can conclude that the 8n variables \quattrova~span the cotangent bundle to the
reversed-parity tangent bundle  which is indicated as~$T^{\star}(\Pi T{\cal M})$.
For more details about this we refer the interested reader to the sixth
paper contained in ref.~\scite{enniocl}.
In the remaining part of this section we will show how to reproduce all the 
abstract Cartan calculus
via our {\it EPB} and the Grassmannian variables. Let us first introduce five 
charges which are
conserved under the $\widetilde{\cal H}$ of eq. (\ref{eq:nove}) and which 
will play an important role in the Cartan calculus. They are:
\bea
\label{eq:undici}
Q^{\scriptscriptstyle BRS}  \equiv  i c^{a}\lambda_{a}~&,&~{\bar
Q}^{\scriptscriptstyle BRS}  \equiv  i {\bar
c}_{a}\omega^{ab}\lambda_{b} \\
\label{eq:dodici}
Q_{g} & \equiv & c^{a}{\bar c}_{a} \\
\label{eq:tredici}
K  \equiv  {1\over 2}\omega_{ab}c^{a}c^{b}~&,&~~{\bar K}\equiv
{1\over 2}\omega^{ab}{\bar c}_{a}{\bar c}_{b}
\eea  

\noindent where $\omega_{ab}$ are the matrix elements of the inverse of 
$\omega^{ab}$.
Next we should note, from the eqs. of motion (\ref{eq:otto}),
that ${\bar c}_{a}$
transform under the Hamiltonian flow as the basis
$\displaystyle\frac{\partial}
{\partial \varphi^{a}}$ of  vector fields. This happens also under any
diffeomorphism of ${\cal M} $. Now since $c^{a}$ transform as basis of forms $d\varphi^{a}$ and 
$\bar{c}_a$ 
as basis of vector fields $\frac{\partial}
{\partial \varphi^{a}}$, let us start building the following map, called
\scite{enniocl} "hat" map :
\bea
\label{eq:quattordici}
\alpha=\alpha_{a}d\varphi ^{a} & \hat{\longrightarrow} &  {\widehat\alpha}\equiv
\alpha_{a}c^{a}\\
\label{eq:quindici}
V=V^{a}\partial_{a}  & \hat{\longrightarrow} & {\widehat V}\equiv V^{a}{\bar
c}_{a}
\eea  
\noindent It is  actually a much more general map between forms $\alpha$, 
antisymmetric
tensors $V$ 
and functions of $\varphi, c, \bar{c}$:
\begin{eqnarray}
\label{eq:sedici}
F^{(p)}={1\over p !}&F_{a_{1}\cdots a_{p}}&d\varphi ^{a_{1}}\wedge\cdots\wedge
d\varphi ^{a_{p}}   \hat{\longrightarrow}\\
&\hat{\longrightarrow}& {\widehat F}^{(p)}\equiv {1\over p!}
F_{a_{1}\cdots a_{p}}c^{a_{1}}\cdots c^{a_{p}}\nonumber
\eea
\bea
\label{eq:diciasette}
V^{(p)}={1\over p!}&V^{a_{1}\cdots a_{p}}&\partial_{a_{1}}\wedge\cdots\wedge 
\partial_{a_{p}}  \hat{\longrightarrow}\\
&\hat{\longrightarrow}&  {\widehat V}\equiv {1\over 
p!}V^{a_{1}
\cdots a_{p}}{\bar c}_{a_{1}}\cdots {\bar c}_{a_{p}}\nonumber
\end{eqnarray} 
\noindent Once the correspondence (\ref{eq:sedici}-\ref{eq:diciasette}) is 
extablished 
we can easily find  out what corresponds in our formalism to the various 
operations of the so called {\it Cartan calculus}\scite{marsd}. They  are
for example the exterior derivative {\bf d}
of a form, or the interior contraction between a vector field $V$ and a form $F$ and
other similar operations. It
is easy\scite{enniocl} to check that:
\begin{eqnarray}
\label{eq:diciotto}
dF^{(p)} & \hat{\longrightarrow} & i\{Q^{\scriptscriptstyle BRS},{\widehat 
F}^{(p)}\}_{\scriptscriptstyle EPB} \\
\label{eq:diciannove}
\iota_{{\scriptscriptstyle V}}F^{(p)} & \hat{\longrightarrow} & i\{{\widehat V},
{\widehat F}^{(p)}\}_{\scriptscriptstyle EPB}
\\
\label{eq:venti}
pF^{(p)} & \hat{\longrightarrow} & i\{Q_{g}, {\widehat 
F}^{(p)}\}_{\scriptscriptstyle EPB}
\end{eqnarray}  
\noindent where $Q^{\scriptscriptstyle BRS}, \, Q_g$ are the charges of 
(\ref{eq:undici}-\ref{eq:dodici}).
In a similar manner we can implement in our language the usual 
mapping~\scite{marsd}
between vector fields $V$
and forms $V^{\flat}$ realized by the symplectic 2-form $\omega(V,0)\equiv 
V^{\flat}$,
or the inverse operation of building a vector field $\alpha^{\sharp}$ out of a 
form:
$\alpha=(\alpha^{\sharp})^{\flat}$. These operations can be turned in our
formalism as follows:
\be
\label{eq:ventitre}
V^{\flat}~~\hat{\longrightarrow}~~ i\{K,{\widehat V}\}_{\scriptscriptstyle 
EPB}~~;~~
\alpha^{\sharp}~~\hat{\longrightarrow}~~i\{{\bar
K},{\widehat\alpha}\}_{\scriptscriptstyle EPB}
\ee  
\noindent where again $K, \bar{K}$ are the charges 
(\ref{eq:tredici}).
We can also implement in our formalism the standard operation\scite{marsd} of building 
a vector field out of a function 
$f(\varphi)$. It is:  
$(df)^{\sharp}  \hat{\longrightarrow}  i\{{\bar Q}^{\scriptscriptstyle 
BRS},f\}_{\scriptscriptstyle EPB}$ 
 \noindent The next thing to do is to reproduce in our formalism the concept of 
Lie derivative\scite{marsd} 
which  is defined as:  
~${\cal L}_{\scriptscriptstyle V}=d\iota_{\scriptscriptstyle V}
+\iota_{\scriptscriptstyle V}d$. It is easy to prove that
\be
{\cal L}_{\scriptscriptstyle V}F^{(p)} \;\; \hat{\longrightarrow} \;\; 
\{-{\widetilde {\cal H}}_{\scriptscriptstyle V},{\widehat F}
^{(p)}\}_{\scriptscriptstyle EPB}
\ee 
\noindent where ${\widetilde {\cal H}}_{\scriptscriptstyle 
V}=\lambda_aV^{a}+i\bar{c}_a
\partial_bV^{a}c^{b}$; note that,
with $V^{a}=\omega^{ab}\partial_bH$,  ${\widetilde {\cal H}}_{\scriptscriptstyle 
V}$ becomes the one
of eq.(\ref{eq:nove}). This confirms that the full ${\widetilde {\cal H}}$ of eq.
(\ref{eq:nove})
is the Lie derivative of the Hamiltonian flow. Finally
the Lie brackets between two vector fields $V,\;W$ are reproduced as:
$[V,W]_{Lie-brack.} \; \hat{\rightarrow} \; \{-{\widetilde {\cal 
H}}_{\scriptscriptstyle 
V},{\widehat W}\}_{\scriptscriptstyle EPB}$. 
In the literature the mathematicians have introduced generalizations
of the Lie-brackets\scite{Kolar}. These are brackets which act on 
form-valued tensor fields and on other similar objects.
Also these brackets can be written in our formalism, for details see the
seventh paper quoted in ref.\scite{enniocl}.
>From all the previous formalism it is now easy to understand why the charges 
of eq.(\ref{eq:undici}) commute
with the Hamiltonian. The first charge is just the exterior derivative,
(see eq.(17)), and it is known from differential geometry
\scite{marsd} that the exterior derivative commutes with the
Lie-derivative. We have called the charges of eq.(\ref{eq:undici}) as BRS charges  because
they anticommute among themselves like the BRS charges of gauge theories do
and moreover, like these ones, they are basically exterior derivatives on some
particular spaces. The five charges of eqs.(\ref{eq:undici}-\ref{eq:tredici})
are not the only ones universally conserved under our
Hamiltonian~${\widetilde{\cal H}}$.
There are also two other ones which are:
$N_{\scriptscriptstyle H}=c^a\partial_aH$ and $\overline{N}_{\scriptscriptstyle
H}=\bar{c}_a\omega^{ab}\partial_ bH$. They can be combined with the
BRS and antiBRS charges in the following manner:
$ Q_{\scriptscriptstyle (1)}\equiv Q^{\scriptscriptstyle BRS}-
\overline{N}_{\scriptscriptstyle H}~~;~~
 Q_{\scriptscriptstyle (2)}\equiv \overline{Q}^{\scriptscriptstyle BRS}+
 N_{\scriptscriptstyle H}$. It is easy to prove that, once the charges and 
 the Hamiltonian
are turned into operators, we get :\break 
$Q_{\scriptscriptstyle (1)}^2=Q^2_{\scriptscriptstyle (2)}=-i{\cal
\widetilde{H}}$.
This means that these charges implement a universal N=2 supersymmetry.
As the $Q^{\scriptscriptstyle BRS}$ of (\ref{eq:undici}) was basically the exterior derivative
on phase space, it would be nice to 
understand the
geometrical meaning also of the susy charges like $Q_{\scriptscriptstyle (1)}$
or $Q_{\scriptscriptstyle (2)}$. 
This was done in the last of references\scite{enniocl}. The strategy used there
was to make  local the global susy invariance  and to analyze the
associated physical state condition. Once this
physical state condition is turned, via the "hat" map of eqs.(13)-(21), into
a Cartan calculus sort of operation, it tells us that the {\it physical} states
are in one to one correspondence with the states of the so called {\it
equivariant cohomology}\scite{Cart} associated to the Hamiltonian vector field.
The equivariant cohomology w.r.t. a vector field V is defined as the set of
forms $|\rho\rangle$ which
satisfy the following conditions:
\bea
(d - \iota_{\scriptscriptstyle V})|\rho\rangle  = 0~~~&;&~~~
{\cal L}_{\scriptscriptstyle V}|\rho\rangle = 0 \nonumber\\
|\rho\rangle \neq (d - \iota_{\scriptscriptstyle V})
|\chi\rangle~~&;&~~{\cal L}_{\scriptscriptstyle V}|\chi\rangle = 0 
\eea
This is basically the geometrical light we could throw on the susy charge
$Q_{\scriptscriptstyle (1)}$. More details can be found in the last paper
of ref.\scite{enniocl}. We found amazing that out of a simple Dirac delta,
like the one in  (\ref{eq:due}),
we have managed to extract all these geometrical structures.
Actually, if we impose  proper boundary conditions, this kind of path-integral 
becomes very similar to the one
of Topological Field Theories\scite{witte}. It is  in fact easy to prove that it
helps in calculating various topological invariants associated
to the phase-space of the system. This was actually done in the fifth paper
of ref.\scite{enniocl}.
All this formalism can be generalized to YM theories\scite{danilo}
and there it may help in studying the geometrical features of the space
of gauge orbits. 
\section{Acknowledgments}
The people I should thank are too many to be listed in a 4-page
brief report. This work was supported by grants from MURST and INFN.

%%%%%%%%%%%%%%%%%%%%%%%%%%%%%%%%%%%%%%%%%%%%%%%%%%%%%%%
%%%%%%%%%%%%%%%%%%%%%%%%%


\begin{thebibliography}{9}


\bibitem{wood1}  N.Woodhouse, "{\it  Geometric quantization}", Claredon Press, Oxford,
1980.

\bibitem{lesha} A.A.Abrikosov (jr) and E.Gozzi, Nucl. Phys. B (Proc.Suppl) vol.88
(2000) 360 (quant-ph-9912050); E.Gozzi et al., work in progress.


\bibitem{enniocl}E.Gozzi, M.Reuter, W.D.Thacker, Phys. Rev. D 40 (1989) 3363;
ibid.D 46, 2 (1992) 757; A.A.Abrikosov (Jr.), Nucl. Phys.
B 382 (1992) 581; E.Gozzi, M.Reuter, Phys. Lett. B 233 (3,4) (1989) 383;
Phys. Lett. B 240 (1,2) (1990) 137; E.Gozzi, M.Regini, Phys.Rev. D 62 (2000)
067702; E.Gozzi, D.Mauro,  Jour. Math. Phys. Vol. 41 no. 4 (200)
1916 ; E. Deotto, E.Gozzi, Int. Jour. Mod. Phys. A Vol. 16 (2001)
2709.

\bibitem{koop}B.O.Koopman, Proc. Nat. Acad. Sci. USA 17 (1931)315; J. von Neumann
Ann. Math. 33 (1932) 587.

\bibitem{marsd}R.Abraham and J.Marsden, "{\it Foundations of Mechanics}",
Benjamin, New York, 1978.

\bibitem{Schw}
A.Schwarz, in "{\it Topics in Statistical and Theoretical Physics}" ed.
R.L.Dobrushin, R.A.Minlos, M.A.Shubin and M.Vershik (AMS, Providence, RI, 1996).
\bibitem{Kolar}
I.Kolar, P.W.Michor and J.Slovak, "{\it Natural Operations in Differential
Geometry}" Springer-Verlag 1993.
\bibitem{Cart}
H. Cartan {\it "Colloque de Topologie"} (Espace Fibres), CBRM 15.71 
Brussels (1950).
\bibitem{witte}
E.Witten, Comm. Math. Phys. 117 (1988) 353.
\bibitem{danilo}
P.Carta, D.Mauro, paper in this proceedings.
\end{thebibliography}
\end{document}